\documentclass[aps,prl,twocolumn,superscriptaddress,showpacs]{revtex4}
\usepackage[dvips]{graphicx}
\usepackage{amsmath}

\renewcommand{\bar}[1]{\overline{#1}}
\newcommand{\half}{{\frac{1}{2}}}

\renewcommand{\bar}[1]{\overline{#1}}

\newcommand{\ket}[1]{\vert\,{#1}\rangle}

\def\Dslash{\raise.15ex\hbox{/}\kern-.7em D}
\def\Pslash{\raise.15ex\hbox{/}\kern-.7em P}

\begin{document}

\preprint{SLAC-PUB-11716}

\title{Hadronic Spectra and Light-Front Wavefunctions in Holographic QCD}

\author{Stanley J. Brodsky}
\affiliation{Stanford Linear Accelerator Center, Stanford University,
Stanford, California 94309, USA}

\author{Guy F. de T\'eramond}
\affiliation{Universidad de Costa Rica, San Jos\'e, Costa Rica}


\begin{abstract}

We show how the string amplitude $\Phi(z)$ defined on the fifth dimension in AdS$_5$ space 
can be precisely mapped to the  light-front wavefunctions of hadrons in physical 
spacetime. We find an exact correspondence between the holographic variable $z$   
and an impact variable $\zeta$, which represents the measure of transverse 
separation of the constituents within the hadrons. In addition, we derive effective 
four dimensional Schr\"odinger equations for the bound states of massless quarks and 
gluons which exactly reproduce the AdS/CFT results and give a realistic
description of the light-quark meson and baryon spectrum as well as the form factors for 
spacelike $Q^2$. Only one parameter which sets the mass scale, $\Lambda_{QCD}$, is introduced.

\end{abstract}

\pacs{11.15.Tk, 11.25Tq, 12.38Aw, 12.40Yx}

\maketitle

The correspondence ~\cite{Maldacena:1997re} between
10-dimensional string theory defined on AdS$_5 \times S^5$
and conformal Yang-Mills gauge theories in physical  3+1
space-time has led to important insights
into the properties of conformal theories at strong coupling.  
QCD is  nearly conformal  in the ultraviolet region. It is  also a
confining gauge theory in the infrared with a mass gap characterized by a scale $\Lambda_{QCD}$
and a well-defined spectrum of color-singlet hadronic
states. Although QCD is not conformal, many aspects of the theory, such as the dimensional 
scaling of exclusive amplitudes~\cite{Brodsky:1973kr}, follow if the QCD coupling has an 
infrared fixed point, 
allowing one to take conformal symmetry as an initial approximation. 

The essential principle which leads to AdS/CFT duality is the fact that the group $SO(2,4)$ 
of Lorentz and conformal transformations has a mathematical representation on AdS$_5$:  
the isomorphism of the
group SO(2,4) of conformal QCD in the limit of
massless quarks and vanishing $\beta$-function
with the isometries of AdS space, $x^\mu \to \lambda x^\mu$, $z \to \lambda z$,
maps scale transformations into the the holographic coordinate $z$,
the extension
of the  hadron wavefunction into the fifth dimension.
Different values of $z$ 
determine the scale of the invariant
separation between quarks.
In particular, the $z \to 0$ boundary corresponds to the $Q \to
\infty$, zero separation limit.
As shown by Polchinski
and Strassler~\cite{Polchinski:2001tt}, the resulting hadronic theory
has the hard behavior and dimensional counting rules~\cite{Brodsky:1973kr} expected 
from a conformal approximation to QCD, rather than the soft
behavior characteristic of string theory.

Color confinement implies that there is a maximum separation of quarks
and a maximum value of $z$.  
The cutoff at $z_0 = 1 /\Lambda_{\rm QCD}$ breaks conformal invariance
and allows the introduction of the QCD scale.
In fact, this holographic model gives a realistic 
description of the light-quark meson and baryon spectrum~\cite{deTeramond:2005su}, 
including orbital 
excitations, as well as the meson and baryon form factors for spacelike $Q^2.$  Remarkably, 
only one parameter $\Lambda_{QCD}$,  enters the predictions.
Essential features  of QCD, its near-conformal behavior at short physical distances
plus color confinement at large interquark separation, are incorporated in the model.
This approach known as holographic QCD,
has been successful in obtaining general
properties of the low-lying hadron spectra, chiral symmetry
breaking, and hadron couplings~\cite{deTeramond:2005su,Csaki:1998qr}.

The light-front wavefunctions $\psi^{S_z}_{n/h}(x_i,
\mathbf{k}_{\perp i},\lambda_i)$ of a hadron $h$ encode all of its bound-state quark and gluon
properties, including its momentum, spin, and flavor
correlations, in the form of universal process- and frame-independent
amplitudes.  In this paper we shall show how the string amplitude $\Phi(z)$ defined on the 
fifth dimension in AdS$_5$ space can be precisely mapped to the boost-invariant light-front 
wavefunctions  of the hadrons.  
The resulting nonperturbative light-front  wavefunctions and distributions allow the calculation 
of many observables including structure functions, distribution amplitudes, form factors, deeply 
virtual Compton scattering and decay constants. 
For example, the scale dependence of the string modes as determined from its twist dimension
as one approaches the $z\to 0$ boundary determines the  power-law behavior
of the  hadronic wavefunction at short distances,  thus providing a precise counting rule
for each Fock component with
any  number of quarks and gluons and any internal orbital
angular momentum~\cite{Brodsky:2003px}.
The predicted short-distance and orbital dependence coincides with perturbative
QCD results~\cite{Ji:2003fw}.

More generally, we shall show that there is an exact correspondence 
between the fifth dimensional 
variable $z$  and a weighted impact separation variable $ \zeta$ in physical space-time 
for each $n$-parton Fock state. Thus the coordinate $z$ can be directly interpreted as a measure 
of the transverse separation of 
the constituents defined by the boost invariant light-front wavefunction (LFWF) 
of the hadronic Fock state.  In addition, we shall derive 
effective radial Schr\"odinger equations for the bound states of massless quarks 
and gluons where the 
effective potential is dictated by conformal symmetry and the constraint that the twist dimension 
of each hadron, including its orbital angular momentum, is reproduced at short distances. These 
effective equations for meson, baryons, and glueballs are completely equivalent to the AdS 
results.

The light-front Fock expansion of any hadronic 
system is constructed by quantizing 
quantum chromodynamics (QCD) at fixed light-cone time 
$x^+ = x^0 + x^3$ and forming the
invariant light-cone Hamiltonian $H_{LC}$: $ H_{LC} = P^2 = P^+ P^-- \mathbf{P}_\perp^2$, 
with $P = (P^+,P^-,\mathbf{P}_\perp)$~\cite{Dirac:1949cp,Brodsky:1997de}.  
In principle, solving the 
$H_{LC}$ eigenvalue problem gives the entire mass spectrum of the
color-singlet hadron states in QCD, together with their respective
light-front wave functions.  A hadronic state satisfies
$H_{LC} \ket{\psi_h} = M^2 \ket{\psi_h}$, 
where $\ket{\psi_h}$ is an expansion in multi-particle Fock eigenstates
$\{\ket{n} \}$ of the free light-front
Hamiltonian: 
$\vert \psi_h \rangle = \sum_n \psi_{n/h} \vert \psi_h \rangle $.
The solutions are
independent of $P^+$ and ${\mathbf{P}_\perp}$. Thus, given the
Fock projections $\psi^{S_z}_{n/h}(x_i, {\mathbf{k}_{\perp i}},\lambda_i)$$
= \langle n;\ x_i, {\mathbf{k}_{\perp i}},
 \lambda_i |\psi_h (P^+, \mathbf{P}_\perp, S_z) \rangle$,
the wave function of a hadron is determined in any frame
\cite{Lepage:1980fj}.
The light-cone momentum fractions $x_i = k^+_i/P^+$ and 
${\mathbf{k}_{\perp i}}$ represent the relative momentum coordinates of 
constituent $i$ in Fock state $n$, and $\lambda_i$ the helicity along
the $\mathbf{z}$ axis. The physical momentum coordinates are
$k^+_i$ and ${\mathbf{p}_{\perp i}}
= x_i {\mathbf{P}_\perp} + {\mathbf{k}_{\perp i}}.$ 
Here
$\sum_{i=1}^n x_i =1$ and 
$\sum_{i=1}^n \mathbf{k}_{\perp i}=0 .$

It is useful to define transverse position coordinates
$x_i \mathbf{r}_{\perp i} = x_i \mathbf{R}_\perp + \mathbf{b}_{\perp i}$
so that $\sum_{i=1}^n \mathbf{b}_{\perp i}=0$ and $ 
\sum_{i=1}^n x_i \mathbf{r}_{\perp i} = \mathbf{R}_\perp$.
The internal coordinates $\mathbf{b}_{\perp i}$ are conjugate to the
relative coordinates $\mathbf{k}_{\perp i}$.
The LFWF $\psi_n(x_j, \mathbf{k}_{\perp j})$ can be expanded in terms of the $n-1$ independent
coordinates $\mathbf{b}_{\perp j}$,  $j = 1,2,\dots,n-1$
\begin{multline}
\psi_n(x_j, \mathbf{k}_{\perp j}) =  (4 \pi)^\frac{n-1}{2} \\
\prod_{j=1}^{n-1}\int d^2 \mathbf{b}_{\perp j}
\exp{\Big(i \sum_{j=1}^{n-1} \mathbf{b}_{\perp j} \cdot \mathbf{k}_{\perp j}\Big)}
\widetilde{\psi}_n(x_j, \mathbf{b}_{\perp j}).
\end{multline}
The normalization is defined by
\begin{equation}  
\sum_n  \prod_{j=1}^{n-1} \int d x_j d^2 \mathbf{b}_{\perp j} 
\vert\widetilde \psi_n(x_j, \mathbf{b}_{\perp j})\vert^2 = 1.
\end{equation}

The electromagnetic current
$J^\mu(0)$  is represented in the interaction picture as a bilinear
product of free fields, so that
it has an elementary coupling to the charged
constituent fields~\cite{Drell:1969km}.
The Drell-Yan-West result for the form factor of a meson 
in terms of the transverse variables $\mathbf{b}_{\perp i}$ has the convenient form:
\begin{multline} \label{eq:FFb} 
F(q^2) =  \sum_n  \prod_{j=1}^{n-1}\int d x_j d^2 \mathbf{b}_{\perp j} \\
\exp{\big(i \mathbf{q}_\perp \cdot \sum_{j=1}^{n-1} x_j \mathbf{b}_{\perp j}\big)}
\vert \widetilde \psi_n(x_j, \mathbf{b}_{\perp j})\vert^2,
\end{multline}
corresponding to a change of transverse momentum $x_j \mathbf{q}_\perp$ for each
of the $n-1$ spectators.  The formula is exact if the sum is
over all Fock states $n$.
We use the standard light-cone frame where
$q = (0, -q^2/P^+,
\mathbf{q}_{\perp})$ and
$P = (P^+, M^2/P^+, \mathbf{0}_{\perp})$. The
momentum transferred by the photon to the system is
$q^2=-2 P \cdot q= -\mathbf{q}_{\perp}^2$. 

The form factor can be related to an effective single particle transverse 
density~\cite{Soper:1976jc}
\begin{equation} \label{eq:FFeta}
F(q^2) =\int^1_0 dx  \int d^2 \vec \eta_\perp e^{i \vec \eta_\perp \cdot \vec q_\perp}
\tilde\rho(x,\vec \eta_\perp).
\end{equation}
From (\ref{eq:FFb}) we find
\begin{multline} \label{eq:rhoeta}
\tilde\rho(x,\vec \eta_\perp) = 
\sum_n \prod_{j=1}^{n-1} \int dx_j d^2\mathbf{b}_{\perp j} 
~\delta \big(1-x-\sum_{j=1}^{n-1} x_j\big) \\
\delta^{(2)}\big(\sum_{j=1}^{n-1} x_j \mathbf{b}_{\perp j} - \vec \eta_\perp\big)
\left\vert \widetilde\psi_n(x_j, \mathbf{b}_{\perp j}) \right\vert^2,
\end{multline}
where the integration is over the coordinates of the $n-1$ spectator partons,
and  $x = x_n$ is the coordinate of the active charged quark.
We can identify
$\vec \eta_\perp = \sum^{n-1}_{j=1} x_j \mathbf{b}_{\perp j}$.
This is the $x$-weighted transverse position coordinate of the $n-1$
spectators.  The procedure is valid for any $n$ and thus the results can be summed over 
$n$ to obtain an exact representation.


We now derive the corresponding expression for the form factor in
AdS. The derivation can
be extended to vector mesons and baryons.
A non-conformal metric
dual to a confining gauge theory is written as~\cite{Polchinski:2001tt}
\begin{equation}
ds^2 = \frac{R^2}{z^2} e^{2A(z)} \left(\eta_{\mu \nu} dx^\mu dx^\nu - dz^2\right),
\label{eq:zmetric}
\end{equation}
where $A(z) \to 0$ as $z \to 0$, and $R$ is the AdS radius.
In the ``hard wall'' approximation~\cite{Polchinski:2001tt}
the non-conformal factor $e^{2A(z)}$ is s a step function: 
$e^{2A(z)} = \theta\big(z \leq \Lambda_{\rm QCD}^{-1}\big)$.

The hadronic matrix element for the electromagnetic current in the 
warped metric (\ref{eq:zmetric}) 
has the form~\cite{Polchinski:2002jw}
\begin{equation}
i g_5 \int d^4x~dz~\sqrt{g}~ A^{\ell}(x,z)
\Phi^*_{P'}(x,z) \overleftrightarrow\partial_\ell \Phi_P(x,z).
\label{eq:M}
\end{equation} 
We take an electromagnetic probe polarized along Minkowski coordinates,
$A_\mu = \epsilon_\mu e^{-i Q \cdot x} J(Q,z)$, $A_z = 0$,
where the function $J(Q,z)$  has the value 1 at
zero momentum transfer, and
as boundary limit the external current 
$A_\mu(x, z \to 0) = \epsilon_\mu e^{-i Q \cdot x}$.
Thus $J(Q=0,z) = J(Q,z=0) = 1$,
since we are normalizing the bulk solutions to the total charge operator.
The solution
to the AdS wave equation, subject to  boundary conditions at  $Q = 0$ and
$z \to 0,$ is~\cite{Polchinski:2002jw}
\begin{equation} \label{eq:J}
J(Q,z) = z Q K_1(z Q).
\end{equation}
The hadronic string modes are plane waves along the Poincar\'e coordinates with four-momentum
$P^\mu$ and invariant mass $P_\mu P^\mu = \mathcal{M}^2$:
$\Phi(x,z) = e^{-iP \cdot x} f(z)$.
Substituting in (\ref{eq:M}) we find
\begin{equation} 
F(Q^2)
= R^{3} \int_0^\infty \frac{dz}{z^{3}}
 e^{3 A(z)} \Phi_{P'}(z) J(Q,z) \Phi_P(z).
\label{eq:FFAdS}
\end{equation}
The form factor in AdS is thus the overlap of the
normalizable modes dual to the incoming
and outgoing hadrons $\Phi_P$ and $\Phi_{P'}$ with the
non-normalizable mode $J(Q,z)$ dual to the external source~\cite{Polchinski:2002jw}.

It is useful to integrate (\ref{eq:FFeta}) over angles to obtain
\begin{equation} \label{eq:FFzeta} 
F(q^2) = 2 \pi \int_0^1 dx \frac{(1-x)}{x}  \int \zeta d \zeta
J_0\left(\zeta q \sqrt{\frac{1-x}{x}}\right) \tilde \rho(x,\zeta),
\end{equation}
where we have introduced the variable
\begin{equation}
\zeta = \sqrt{\frac{x}{1-x}} ~\Big\vert \sum_{j=1}^{n-1} x_j \mathbf{b}_{\perp j}\Big\vert,
\end{equation}
representing the $x$-weighted transverse impact coordinate of the spectator system.
We also note the identity
\begin{equation} \label{eq:intJ}
\int_0^1 dx J_0\left(\zeta Q \sqrt{\frac{1-x}{x}}\right) = \zeta Q K_1(\zeta Q),
\end{equation}
which is precisely the solution  $J(Q,\zeta)$
for the electromagnetic potential in AdS (\ref{eq:J}).
We can now see the equivalence between the LF and AdS results for the hadronic form factors.
Comparing (\ref{eq:FFzeta}) with the expression for the form factor in
AdS space (\ref{eq:FFAdS}), we can identify the spectator density
function appearing in the light-front 
formalism with the corresponding AdS density
\begin{equation} \label{eq:hc}
\tilde \rho(x,\zeta)
 =  \frac{R^3}{2 \pi} \frac{x}{1-x} e^{3 A(\zeta)}
\frac{\left\vert \Phi(\zeta)\right\vert^2}{\zeta^4}.
\end{equation}
Equation (\ref{eq:hc}) gives a precise relation between  string modes $\Phi(\zeta)$ 
in AdS$_5$ and the QCD transverse charge density $\tilde\rho(x,\zeta)$. 
The variable $\zeta$, $0 \le \zeta \le \Lambda^{-1}_{\rm QCD}$, represents the invariant
separation between point-like constituents, and it is also the
holographic variable $z$ 
characterizing the string scale in AdS; {\it i.e.}, we can identify $\zeta = z$.
For example, for two partons
$\tilde\rho_{n=2}(x, \zeta) =
\vert \psi(x, \zeta)\vert^2/(1-x)^2$,
and a closed form solution for the two-constituent bound state
light-front wave function is found
\begin{equation} \label{eq:Phipsi}
\left\vert\widetilde\psi(x,\zeta)\right\vert^2 =
\frac{R^3}{2 \pi} ~x(1-x)~ e^{3 A(\zeta)}~
\frac{\left\vert \Phi(\zeta)\right\vert^2}{\zeta^4}.
\end{equation}
In the case of two partons 
$\zeta^2 = \frac{x}{1-x} \vec\eta_\perp^2
      = x(1-x) \mathbf{b}_\perp^2$.

\begin{figure}[h]
\centering
\includegraphics[width=8.2cm]{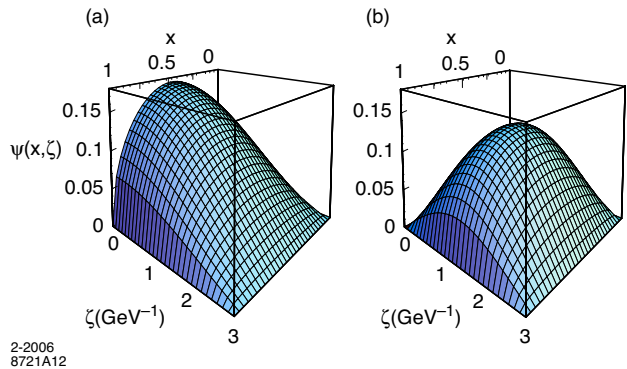}
\label{fig:AdSLFWF}
\caption{\small Two-parton bound state holographic LFWF $\widetilde\psi(x, \zeta)$ for
$\Lambda_{\rm QCD} = 0.32$ GeV:
(a) ground state $L = 0$, $k = 1$, (b) first orbital excited
state $L = 1$, $k = 1$.}
\label{fig:LFWF}
\end{figure}

\begin{figure}[h]
\centering
\includegraphics[angle=0,width=8.2cm]{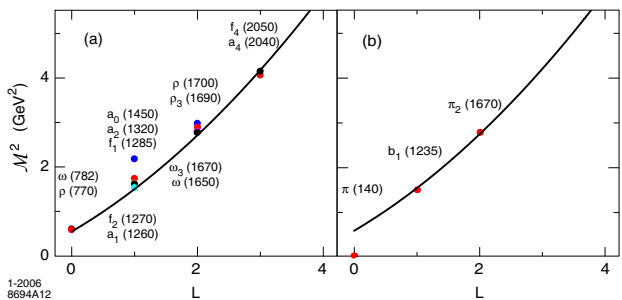}
\caption{Light meson orbital states for $\Lambda_{\rm QCD}$ = 0.32 GeV:
(a) vector mesons and (b) pseudoscalar mesons.}
\label{fig:MesonSpec}
\end{figure}

In general, the short-distance behavior of a hadronic state is
characterized by its twist  (dimension minus spin) 
$\tau = \Delta - \sigma$, where $\sigma$ is the sum over the constituent's spin
$\sigma = \sum_{i = 1}^n \sigma_i$. Twist is also equal to the number of partons $\tau = n$.
Upon the substitution $\Delta \to n + L$, $\phi(z) = z^{-3/2} \Phi(z)$ in the  
AdS wave equations
describing glueballs, mesons or vector mesons~\cite{deTeramond:2005su},
we find an effective Schr\"odinger equation as a function of the
weighted impact variable $\zeta$
\begin{equation} \label{eq:Scheq} 
\left[-\frac{d^2}{d \zeta^2} + V(\zeta) \right] \phi(\zeta) = \mathcal{M}^2 \phi(\zeta),
\end{equation}
with the effective conformal potential~\cite{Rey:1998ik}
\begin{equation}
V(\zeta) = - \frac{1-4 L^2}{4\zeta^2}.
\end{equation}
This new effective LF wave equation in physical space-time has 
stable solutions satisfying the 
Breitenlohner-Freedman bound~\cite{Breitenlohner:1982jf}. The solution
to (\ref{eq:Scheq}) is  
\begin{equation}
\phi(z) = z^{-\frac{3}{2}} \Phi(z) = C z^\frac{1}{2} J_L(z \mathcal{M}).
\end{equation}
Its eigenvalues are determined by
the boundary conditions at $\phi(z = 1/\Lambda_{\rm QCD}) = 0$ and
are given in terms of the roots of the Bessel functions: 
$\mathcal{M}_{L,k} = \beta_{L,k} \Lambda_{\rm QCD}$.
Normalized LFWFs $\widetilde\psi_{L,k}$ follow from
(\ref{eq:Phipsi}) ~\cite{Radyushkin:1995pj} 
\begin{multline} 
\widetilde \psi_{L,k}(x, \zeta) 
=  B_{L,k} \sqrt{x(1-x)} 
J_L \left(\zeta \beta_{L,k} \Lambda_{\rm QCD}\right) \\
\theta\big(z \le \Lambda^{-1}_{\rm QCD}\big),
\end{multline}
where $B_{L,k} = {\Lambda_{\rm QCD}}
  \left[(-1)^L \pi J_{1+L}(\beta_{L,k}) J_{1-L}(\beta_{L,k})\right]^{-\half}$.
The first eigenmodes are depicted in Figure \ref{fig:LFWF}, and the masses of the
light mesons in Figure \ref{fig:MesonSpec}. The predictions for the lightest hadrons are improved 
relative to the results of~\cite{deTeramond:2005su} using the  boundary conditions determined
in terms of twist instead of conformal dimensions. The description of baryons is carried out
along similar lines and will be presented elsewhere.

The holographic model is remarkably successful in organizing the hadron spectrum, although
it underestimates the spin-orbit splittings of the $L=1$ states. A better understanding
of the relation between chiral symmetry breaking and confinement is required to
describe successfully the pion. This would probably need a description
of quark spin-flip mechanisms at the wall.

We have shown how the string amplitude $\Phi(z)$ defined on the fifth dimension in $AdS_5$ 
space can be precisely mapped to the frame-independent light-front wavefunctions of hadrons 
in physical spacetime. This specific correspondence provides an exact holographic mapping
at all energy scales between string modes in AdS and boundary states with well-defined number 
of partons. Consequently, the AdS string mode $\Phi(z)$ can be regarded as the probability
amplitude to find $n$-partons at transverse impact separation $\zeta =
z$. Its eigenmodes determine the hadronic mass spectrum. 
The degeneracy of hadron states depends on the flavor symmetry that is
assumed; {\it i.e.},  the number of massless quarks. There is no explicit dependence on $N_C$,
and  the QCD spectrum follows by matching twist dimensions to
$SU(3)_C$ color-singlet hadronic states at the $z\to 0$ boundary.

The model can also be formulated in four dimensions without reference to AdS 
space~\cite{Erlich:2006hq}. To this end
we have derived effective radial Schr\"odinger equations for the bound states of massless quarks 
and gluons with boundary conditions at zero separation distance determined by twist. These 
effective equations for meson, baryons, and glueballs exactly reproduce the AdS/CFT results.  
Since only one parameter is introduced, the 
agreement of the hadron spectrum with the observed pattern of physical states and the behavior 
of measured spacelike form factors is remarkable. When one includes the orbital dependence, the resulting set of wavefunctions are orthonormal and complete, giving a correct representation of current and charge matrix elements.

The phenomenological success of dimensional counting rules for
exclusive processes can be understood if QCD
resembles a strongly coupled conformal theory. The holographic 
model gives a mathematical realization of such theories.  In some sense it is a
covariant generalization of the MIT bag model, but it also incorporates
the approximately conformal behavior of QCD at short physical distances.
Our results suggest that basic features of QCD can be understood in terms
of a higher dimensional dual gravity theory 
which  holographically encodes multi-parton boundary states into string modes and 
allows the computation of physical observables at strong coupling.

\begin{acknowledgments}

This research was supported by the Department
of Energy contract DE--AC02--76SF00515.

\end{acknowledgments}

\end{document}